%% file: paper.tex
\documentclass[letterpaper,twocolumn,10pt]{article}
\usepackage{usenix-2020-09}
\usepackage{tabularx, booktabs, pifont}
\usepackage{graphicx,setspace,listings,multicol,multirow,xspace,enumitem,booktabs,subcaption}
\usepackage{amsmath}
\usepackage{amssymb}
\usepackage{algorithm}
\usepackage{algpseudocode}
\usepackage{comment}
\usepackage{balance}
\usepackage{enumitem}
\usepackage{tikz}
\usetikzlibrary{calc}
\usepackage{pifont}
\usepackage{makecell}

\input{macros}

\date{}

\title{\Large \bf VASTRA: Efficient Agentic Top-K Interactive Video Analysis}

\author{
    \begin{tabular}{c}
        \textbf{Tushar Tyagi} \\
        Georgia Institute of Technology \\
        \tt{ttyagi9@gatech.edu}
    \end{tabular}
    \hfill %
    \begin{tabular}{c}
        \textbf{Aditya Raghavan} \\
        Georgia Institute of Technology \\
        \tt{araghavan68@gatech.edu}
    \end{tabular}
    \hfill %
    \\[2em] %
    \begin{tabular}{c}
        \textbf{Haoran Qiu} \\
        Microsoft Azure Research \\
        \tt{haoranqiu@microsoft.com}
    \end{tabular}
    \qquad\qquad %
    \begin{tabular}{c}
        \textbf{Francisco Romero} \\
        Georgia Institute of Technology \\
        \tt{faromero@gatech.edu}
    \end{tabular}
}

\begin{document}

\maketitle

\input{tex/abstract.tex}

\input{tex/introduction.tex}
\input{tex/motivation.tex}
\input{tex/system.tex}
\input{tex/evaluation.tex}
\input{tex/related_work.tex}
\input{tex/conclusion.tex}

{\footnotesize \bibliographystyle{plain}
\bibliography{paper}}

\end{document}

%% file: macros.tex
\newcommand*{\sn}{\textsc{VASTRA}\xspace}

\setlist[itemize]{leftmargin=*}
\setlist[enumerate]{leftmargin=*}

\newcommand*{\eg}{\emph{e.g.,}~}

\newcommand{\minihead}[1]{{\vspace{.45em}\noindent\textbf{#1.~~}}}

\usepackage{etoolbox}
\newtoggle{showmarks}
\toggletrue{showmarks}

\iftoggle{showmarks}{
  \newcommand\franky[1]{\textcolor{blue}{F: #1}}
  \newcommand\aditya[1]{\textcolor{teal}{A: #1}}
  \newcommand\tushar[1]{\textcolor{violet}{T: #1}}
  \newcommand\haoran[1]{\textcolor{orange}{H: #1}}
}{
  \newcommand\franky[1]{\unskip}
  \newcommand\aditya[1]{\unskip}
  \newcommand\tushar[1]{\unskip}
  \newcommand\haoran[1]{\unskip}
}

%% file: tex/abstract.tex
\subsection*{Abstract}
Vision-language models (VLMs) enable natural language video retrieval, bypassing traditional hand-engineered pipelines. Integrating agentic reasoning frameworks (e.g., ReAct) further advances this paradigm by allowing models to dynamically inspect video content rather than relying on a single visual pass. Applying this pattern naively to video retrieval inherits problems that video systems cannot afford: results vary from run to run under identical conditions, and cost and latency scale with every reasoning step the agent takes rather than with the work the query actually requires. We present \textsc{Vastra}, an agentic video retrieval system whose execution is dictated by query complexity rather than an open-ended reasoning loop: cheap, reliably answerable queries are ranked directly from pre-ingested summaries, while harder queries trigger targeted verification only where relevance is genuinely uncertain. This keeps control flow external to the model rather than latent in a growing chain of thought, establishing deterministic control flow by construction and end-to-end rank reproducibility under greedy serving. Across 40 benchmark queries, \textsc{Vastra} exposes explicit presets: \textsc{Vastra}-Fast runs $12.2\times$ faster ($4.2$\,s vs.\ $51.2$\,s) and $10.4\times$ cheaper ($\$0.0044$ vs.\ $\$0.0458$) than best performing ReAct agent at a 5.0-point mAP reduction, while \textsc{Vastra}-Max matches peak ReAct agent accuracy ($0.653$ vs.\ $0.649$ mAP) at comparable latency, trading higher spend ($\$0.1022$ vs.\ $\$0.0458$) for zero run-to-run variance.

%% file: tex/introduction.tex
\section{Introduction} \label{sec:introduction}

Video analytics is now prevalent across surveillance, retail, transportation, and industrial monitoring as camera deployments proliferate.
Traditional systems~\cite{noscope,blazeit,focus,boggart,videochef,everest,eva,seiden} chain task-specific models into fixed pipelines: accurate for the predicate they were built for, but rigid.
Adding a new query type requires retraining or reconfiguring the pipeline.
No amount of configuration lets these systems answer open-ended, natural-language questions about video content.

Foundation models remove that rigidity.
A single vision-language model (VLM) can answer diverse questions about the same clip, collapsing multi-model pipelines and accepting queries in natural language.
Agentic approaches extend this flexibility: instead of relying on a single forward pass, an agent interleaves reasoning with tool calls, deciding at each step what to inspect next and revising its plan based on intermediate observations~\cite{yao2023react}.
For video, this enables inspecting targeted intervals, gathering context, or changing course mid-query when an initial pass cannot resolve user intent.

That flexibility, however, introduces three challenges that current multimodal agent systems leave unresolved:
\begin{enumerate}[leftmargin=*, label=\textbf{\arabic*.}]
    \item \textbf{Determinism.}
    Atil et al.~\cite{atil2024nondeterminism} show that repeated runs of the same prompt under nominally deterministic settings vary in accuracy by up to 15\%, driven by non-associative floating-point arithmetic and GPU scheduling below the API configuration layer.
    For video retrieval, the same query over the same corpus can silently return a different Top-$K$ ranking each time, breaking cacheability and output reproducibility.
    
    \item \textbf{Listwise degradation on long candidate lists.}
    A retrieval agent must order a pool of candidates, but LLMs struggle when reasoning over an entire pool simultaneously.
    In this listwise formulation, accuracy drops as candidate lists grow: models attend poorly to information in the middle of long inputs~\cite{liu2024lostmiddle}.
    Listwise LLM ranking stays dependable only for small candidate lists, degrading into dropped items beyond them.
    Models judge candidates far more reliably when applied pointwise to score a single candidate summary against the query.
 
    \item \textbf{Cost and latency.}
    Examining video frames across candidates requires expensive model invocations, and agentic reasoning compounds this by re-invoking models on growing contexts at every step.
    Because each action depends sequentially on prior observations, tool latencies accumulate rather than executing in parallel.
    Even bounded candidate rerankers compound latency and spend: each step re-invokes the model on expanding context to dispatch multimodal tool calls sequentially across candidates, averaging tens of seconds per query.
\end{enumerate}

Existing systems address these challenges only partially.
ReAct-style agents~\cite{yao2023react} grant models open-ended inspection autonomy, resulting in unbounded spend and non-deterministic rankings.
Ava~\cite{ava} constructs event knowledge graphs for ultra-long stream question answering, but inherits similar per-query variance and targets single-stream analysis rather than candidate ranking.
Circinus~\cite{circinus} optimizes compound pipelines for latency SLOs across fixed model sets, but does not adapt VLM reasoning to query complexity.
None of these systems size verification effort to query difficulty: they apply either fixed pipelines or unbounded agent deliberation across all candidates.

\smallskip
We present \textsc{Vastra}, an agentic video retrieval system that sizes reasoning to query complexity.
Rather than relying on an open-ended deliberation loop, \textsc{Vastra} compiles queries into explicit directed acyclic predicate graphs and restricts multimodal verification to uncertain candidates.
At ingestion, a VLM summarizes each video once into reusable text descriptions and vector embeddings, amortizing visual processing across all future queries.
At query time, an LLM decomposes the query into independent anchors and dependent constraints, routing the query to summary scoring (\textsc{Route~A}), a sequential cascade (\textsc{Route~B}), or a hybrid join (\textsc{Route~MIXED}) based on graph topology.
For routed candidates, \textsc{Vastra} scores summaries in parallel chunks and partitions candidates into discrete tiers: confident matches resolve immediately, while expensive VLM clip examination is restricted strictly to uncertain candidates in the margin.
Because compiled predicate programs govern control flow rather than sampling in an agent loop, \textsc{Vastra} returns identical rankings on repeated runs under greedy serving while keeping execution spend bounded.

We evaluate \textsc{Vastra} on 40 natural-language queries spanning diverse visual characteristics across two video domains (everyday indoor activities from Charades~\cite{sigurdsson2016hollywood}, and dashcam and web footage), totaling 229 ingested videos (over 2.6 hours of footage), against a suite of ReAct baselines (\S\ref{sec:evaluation}).
\textsc{Vastra}'s presets establish explicit operating trade-offs: \textsc{Vastra}-Fast runs $12.2\times$ faster ($4.2$\,s vs.\ $51.2$\,s) and $10.4\times$ cheaper ($\$0.0044$ vs.\ $\$0.0458$) than multimodal ReAct-S+E at a $5.0$-point mAP drop ($0.599$ vs.\ $0.649$)---and outperforms like-for-like summary-only ReAct-S by $+1.4$ mAP points ($0.599$ vs.\ $0.585$) while running $6.5\times$ faster and costing $2.6\times$ less ($\$0.0044$ vs.\ $\$0.0115$).
Meanwhile, \textsc{Vastra}-Max matches peak baseline accuracy ($0.653$ vs.\ $0.649$ mAP, within sample margin, $p=0.84$) at comparable latency ($48.2$\,s vs.\ $51.2$\,s), trading twice the candidate pool ($K=40$ vs.\ $K=20$) and higher spend ($\$0.1022$ vs.\ $\$0.0458$) for exact run-to-run determinism ($\sigma = 0.000$, $\tau = 1.000$) under greedy serving, where ReAct's rankings exhibit run-to-run standard deviations of $\pm 0.021$ (and up to $\pm 0.037$ under ablated reasoning).

\smallskip
\noindent We make the following contributions:
\begin{enumerate}[leftmargin=*, label=\textbf{\arabic*.}]
    \item A characterization of failure modes when applying open-ended VLM agents to Top-$K$ video retrieval, demonstrating non-determinism, uncontrolled working-set explosion, and unbounded cost.
    \item \textbf{\textsc{Vastra}}, an agentic Top-$K$ video reranking system whose central technique is complexity- and uncertainty-aware allocation of VLM work: predicate decomposition plus routing that examines only uncertain candidates, yielding deterministic, bounded-cost ranking.
    \item An evaluation across two video domains and a suite of ReAct baselines showing \textsc{Vastra} spans orders-of-magnitude faster triage to deterministic high-accuracy reranking, reproduces rankings identically under greedy serving, and maintains accuracy across model families, with an analysis isolating which design decisions produce each gain.
\end{enumerate}

%% file: tex/motivation.tex
\section{Why Interactive Video Search Is Hard}
\label{sec:motivation}

Applying foundation models to video \emph{retrieval} introduces trade-offs absent from single-clip question answering~\cite{blazeit,yao2023react}: reasoning must execute across an entire candidate pool under strict cost and latency constraints, and the resulting ranking must remain stable.
Table~\ref{tab:features} places existing system classes against the properties interactive retrieval requires.
While classical analytics and CLIP-based retrieval provide bounded latency, they lack open-ended natural-language expression or fine-grained visual verification.
Conversely, unconstrained VLM agents offer natural-language flexibility, but fail to guarantee deterministic rankings, bounded costs, or workload scaling.

\begin{table}[t]
\centering
\caption{Comparison of interactive video retrieval architectures across system requirements. Classical = proxy/cascade analytics (NoScope, BlazeIt); CLIP-RAG = embedding retrieval (Zelda); VLM-agent = ReAct-style / Ava. Unlike heuristic cascades, embedding search, or unconstrained VLM agents, \textsc{Vastra} achieves deterministic, complexity-proportional verification under bounded cost. (\ding{51}: yes; $\sim$: partial; \ding{55}: no).}
\label{tab:features}
\footnotesize
\setlength{\tabcolsep}{3pt} %
\begin{tabularx}{\columnwidth}{>{\raggedright\arraybackslash}X cccc}
\toprule
\textbf{Property} & \textbf{Class.} & \textbf{CLIP} & \textbf{VLM} & \textbf{\textsc{Vastra}} \\
\midrule
Natural-language queries          & \ding{55} & \ding{51} & \ding{51} & \ding{51} \\
No per-query retraining           & \ding{55} & \ding{51} & \ding{51} & \ding{51} \\
Targeted visual verification      & $\sim$    & \ding{55} & \ding{51} & \ding{51} \\
Deterministic across runs         & \ding{51} & \ding{51} & \ding{55} & \ding{51} \\
Bounded, predictable cost         & \ding{51} & \ding{51} & \ding{55} & \ding{51} \\
Effort scaled to complexity       & \ding{55} & \ding{55} & \ding{55} & \ding{51} \\
\bottomrule
\end{tabularx}
\end{table}

\subsection{Decomposition Helps Some Queries and Hurts Others}
\label{sec:motivation-decomposition}

A natural strategy to bound agentic verification is to compile natural-language queries into explicit, structured sub-predicates and verify them independently.
To evaluate the impact of this strategy prior to building a full router, we conducted an exploratory study comparing predicate-decomposed execution against monolithic query verification across our benchmark query archetypes using the same VLM backbone (\texttt{gemini-2.5-flash}, greedy decoding $T=0.0$).
For the decomposed baseline, candidate clips are evaluated across sequentially chained predicates with short-circuiting; for the monolithic baseline, the VLM is prompted with the complete, undecomposed query string in a single verification pass.

\begin{figure}[t]
\centering
\makebox[\columnwidth][c]{%
  \includegraphics[width=1.12\columnwidth]{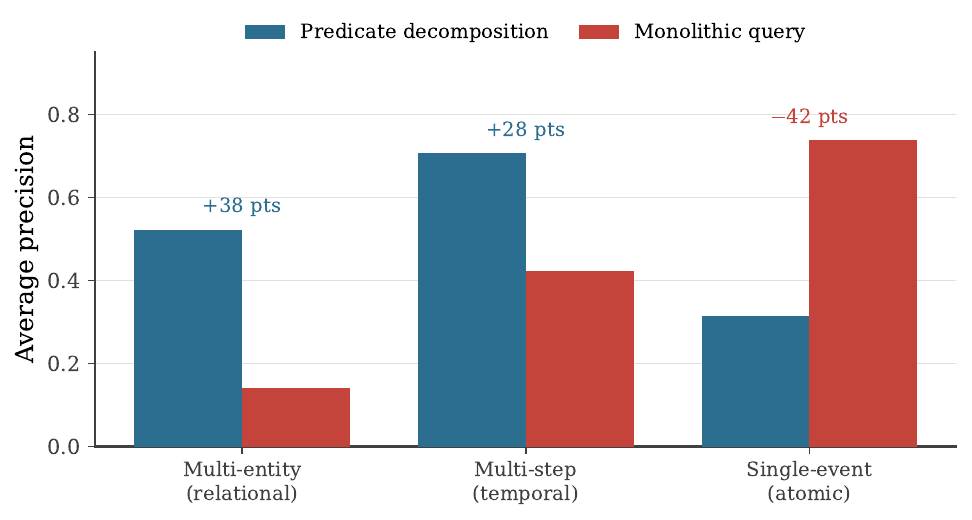}%
}
\caption{Predicate decomposition vs.\ monolithic query verification across query categories ($K=20$, $T=0.0$). Structured decomposition substantially improves multi-entity relational queries (+38 pts) and multi-step temporal sequences (+28 pts), but degrades atomic single-action queries ($-42$ pts), demonstrating that neither strategy dominates globally and motivating topology routing (\S\ref{sec:system}).}
\label{fig:decomp}
\end{figure}

Comparing the two execution paradigms reveals sharp heterogeneity across query types (Figure~\ref{fig:decomp}).
Predicate decomposition improves average precision by $+38$ points on \textbf{multi-entity} (relational) queries ($0.142 \to 0.523$ AP) and $+28$ points on \textbf{multi-step} (temporal) queries ($0.424 \to 0.707$ AP) by isolating distinct entities and ordered sub-events that a monolithic prompt conflates.
Conversely, decomposition \emph{hurts} \textbf{single-event} (atomic) queries by $-42$ points ($0.738 \to 0.315$ AP): splitting an indivisible action into mandatory sub-clauses allows a single false-negative verdict on an incidental detail to demote a true target.

Decomposition is therefore neither universally beneficial nor harmful: it succeeds on structurally complex queries but fails on atomic ones.
Always decomposing degrades atomic queries, while never decomposing degrades complex queries.
This trade-off necessitates query-dependent execution: routing queries whose predicates are answerable from compact summaries directly to text scoring, while reserving multi-stage verification cascades for queries with relational dependencies or summary-unanswerable visual details (\S\ref{sec:system}).

%% file: tex/system.tex
\section{\sn} \label{sec:system}

The goal of \sn is to rank a Top-$K$ candidate pool against natural-language queries while avoiding the cost, latency, and variance of unconstrained agent loops (\S\ref{sec:motivation}).
The system compiles queries into explicit execution graphs of independent \textbf{Anchors} (primary attributes) and conditioned \textbf{Dependents} (secondary spatial, temporal, or acoustic constraints), inspecting video footage only when textual evidence is genuinely uncertain.

\subsection{Predicate Breakdown and Router}
\label{sec:sys-router}

\begin{figure*}[t]
\centering
\includegraphics[width=\textwidth]{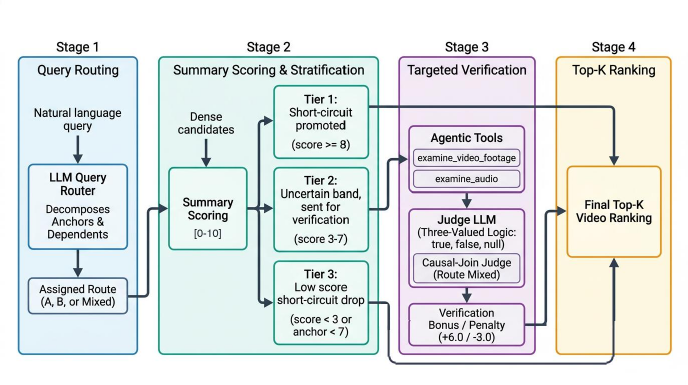}
\caption{The \sn system workflow. Queries are routed based on complexity. For complex queries, initial candidates are partitioned into tiers, restricting expensive multimodal verification to the ``Uncertain Band'' (Tier 2). FAST and MAX modes offer explicit latency/recall trade-offs.}
\label{fig:workflow}
\end{figure*}

The router determines execution strategy dynamically based on predicate properties (Type, Persistence, Dependency), as uniform processing across diverse queries is suboptimal (\S\ref{sec:motivation-decomposition}).
\sn decouples structural execution paths, termed \textbf{Routes} (dictated by query complexity), from user-selected cost-latency profiles, termed \textbf{Modes} (Fast, Normal, and Max).

\paragraph{Predicate Extraction and Complexity Formulation.}
    Given natural-language query $q$, the decision LLM extracts atomic predicates via structured schema generation.
    The extraction prompt decomposes the query into minimal independently verifiable events that preserve actor-action bindings, classifying
  each into an operational taxonomy (\eg visual attributes, temporal sequences, causal joins, exact counts, duration limits, or audio cues).
    For each predicate, the model assigns persistence (persistent vs.\ transient), estimated candidate selectivity, timestamp requirements,
  and parent dependencies (\texttt{depends\_on}).
    These parent references define directed edges $\mathcal{E}_q$, constructing directed acyclic predicate graph $G_q = (\mathcal{P}_q,
  \mathcal{E}_q)$ with vertices $\mathcal{P}_q$.
    Predicates with zero in-degree define anchors $\mathcal{A}_q = \{p \in \mathcal{P}_q \mid \operatorname{in\text{-}degree}(p) = 0\}$,
  while remaining vertices define dependents $\mathcal{D}_q = \mathcal{P}_q \setminus \mathcal{A}_q$.
    Query complexity is characterized by the tuple:
    \begin{equation}
        \mathcal{C}(q) = \langle |\mathcal{P}_q|, \, \operatorname{depth}(G_q), \, \mathbb{I}_{\text{unans}}(q) \rangle
    \end{equation}
    where $\operatorname{depth}(G_q)$ denotes the longest dependency path from root anchor to leaf, and $\mathbb{I}_{\text{unans}}(q) \in \{0,
  1\}$ indicates whether any extracted predicate is transient or requires modalities unanswerable from text summaries alone.

\paragraph{Four Ordered Routing Rules and Execution Routes.}
The router evaluates $\mathcal{C}(q)$ against four ordered decision rules:
\begin{enumerate}
    \item \textbf{Rule 1 (Zero-Verification Gate):} If $q$ contains no verifiable predicates (\eg subjective impressions or broad scene descriptions), assign \textbf{Route A}.
    \item \textbf{Rule 2 (Selectivity Anchor Gate):} If $q$ lacks an objective, high-selectivity anchor to quarantine candidate footage, assign \textbf{Route A}.
    \item \textbf{Rule 3 (Summary-Answerable Gate):} If $\mathbb{I}_{\text{unans}}(q) = 0$ (all predicates can be evaluated from pre-computed text summaries), assign \textbf{Route A}.
    \item \textbf{Rule 4 (Topology Selection):} For queries passing Rules 1--3 (possessing a selective anchor and at least one summary-unanswerable predicate), assign execution routes by graph topology:
    \begin{itemize}
        \item \textbf{Route B (Sequential Cascade):} For sequential chains ($\operatorname{depth}(G_q) \ge 1$), atomic visual predicates ($|\mathcal{P}_q| = 1$), or independent multi-attribute queries ($\operatorname{depth}(G_q) = 0, |\mathcal{P}_q| > 1$), verify predicates sequentially with candidate short-circuiting.
        \item \textbf{Route MIXED (Hybrid Relational Join):} For cross-modal conditions or temporal joins ($\operatorname{depth}(G_q) \ge 1$ with relational edges), extract candidate timestamps across predicates in parallel before executing a joint temporal alignment and causal verification pass.
    \end{itemize}
\end{enumerate}
Route A executes pure summary scoring without multimodal verification.
Routes B and MIXED execute tiered candidate pruning followed by targeted visual or audio examination.

\paragraph{Operating Modes and Presets.}
Independent of the assigned route, \sn executes under one of three user-selected \textbf{Modes}:
\begin{itemize}
    \item \textbf{Fast Mode:} Bypasses downstream VLM verification entirely, ranking candidates solely via parallel text summary scoring.
    \item \textbf{Normal Mode:} Evaluates candidates through selective tiering, triggering VLM verification only for uncertain summaries.
    \item \textbf{Max Mode:} Enforces exhaustive verification on candidates with viable anchors, employing chunk-level judge aggregation and demotion-repair passes.
\end{itemize}
In deployment, \textsc{Vastra} packages these configurations into three presets: \textsc{Vastra}-Fast (Fast Mode, $K=20$), \textsc{Vastra}-Balanced (Normal Mode, $K=20$), and \textsc{Vastra}-Max (Max Mode, $K=40$). K defines the number of top-k candidates being re-ranked by the system.

\subsection{Generating and Validating Candidates} \label{sec:optimization}

Figure~\ref{fig:workflow} outlines candidate generation and validation across offline ingestion and online execution.

 \paragraph{Candidate Ingestion and Dense Retrieval.}
    During offline ingestion, a VLM summarizes each video into a concise text description indexed for dense retrieval with FAISS~\cite{faiss}
  over \texttt{MiniLM} embeddings~\cite{wang2020minilm}.
    At query time, \sn embeds query $q$ to retrieve an initial candidate pool $\mathcal{C} = \{v_1, \dots, v_K\}$ ($K \in \{20, 40\}$ by
  operating mode) via vector similarity over summary embeddings.

    \paragraph{Chunked Summary Scoring Pass.}
    \sn scores candidate summaries against decomposed predicates in parallel batches: the decision LLM evaluates candidates five per prompt
  to prevent context dilution.
    For each candidate video $v \in \mathcal{C}$, this pass produces predicate confidence scores $s(v, p) \in [0, 10]$ for each predicate $p$,
  an overall summary relevance score $S_{\text{summary}}(v) \in [0, 10]$, and an anchor score $s(v, \mathcal{A}_q) = \max_{a \in
  \mathcal{A}_q} s(v, a)$ using three-valued evidence: \texttt{SUPPORTED}, \texttt{UNMENTIONED}, or \texttt{CONTRADICTED}.

\paragraph{Tiered Candidate Partitioning and Mid-Band Filtering.} \label{sec:uncertainty-scoring}
Because lossy text summaries frequently omit brief visual events, running a VLM over every candidate inflates cost and latency.
\sn bounds inspection by defining an uncertainty interval for dependent predicates:
\begin{equation}
    \text{DEP\_BAND} = [\tau_{\text{lo}}, \, \tau_{\text{hi}}] = [3, \, 7]
\end{equation}
For any candidate $v$, its unresolved verification set $\mathcal{U}(v)$ comprises dependent predicates falling within this interval, joined with relational predicates $\mathcal{D}_{\text{join}}$ that summaries cannot settle:
\begin{equation}
    \mathcal{U}(v) = \left\{ d \in \mathcal{D}_q \setminus \mathcal{D}_{\text{join}} \;\middle|\; s(v, d) \in [\tau_{\text{lo}}, \tau_{\text{hi}}] \right\} \cup \mathcal{D}_{\text{join}}
\end{equation}
With candidate uncertainty count $u(v) = |\mathcal{U}(v)|$, candidates are partitioned into three mutually exhaustive tiers:
\begin{itemize}
    \item \textbf{Tier 1 (High Summary Score):} Candidates with $S_{\text{summary}}(v) \ge \tau_{\text{short}}$ ($\tau_{\text{short}} = 8$) bypass multimodal verification and promote directly to top ranks.
    \item \textbf{Tier 2 (Uncertain Verification Set, $\mathcal{V}_{\text{verify}}$):} Candidates with confirmed anchors but unresolved dependent uncertainty:
    \begin{equation}
        \mathcal{V}_{\text{verify}} = \left\{ v \in \mathcal{C} \setminus \text{Tier 1} \;\middle|\; s(v, \mathcal{A}_q) \ge \tau_{\text{anchor}} \;\land\; u(v) \ge 1 \right\}
    \end{equation}
    where $\tau_{\text{anchor}} = 7$. Only candidates in $\mathcal{V}_{\text{verify}}$ trigger VLM tool calls, restricted strictly to predicates in $\mathcal{U}(v)$.
    \item \textbf{Tier 3 (Baseline Remainder, $\mathcal{V}_{\text{rest}}$):} Candidates failing the anchor threshold or whose predicates are already settled ($u(v) = 0$). These bypass visual inspection and rank directly by base summary score.
\end{itemize}
If $\mathcal{V}_{\text{verify}}$ is empty, fallback tiers F1 (verifying dependents for confirmed anchors) or F2 (verifying anchors for top baseline candidates, if no anchor is confirmed) activate to ensure recall.

\paragraph{Targeted Verification and Three-Valued Logic.}
For candidates in $\mathcal{V}_{\text{verify}}$, \sn dispatches targeted tools (\texttt{examine\_video\_footage}, \texttt{examine\_audio}) in parallel across candidates and video chunks.
A Judge LLM evaluates chunk outputs using three-valued logic: \texttt{TRUE} (supported), \texttt{FALSE} (contradicted), or \texttt{NULL} (inconclusive/unmentioned).
For relational conditions, a Causal-Join Judge verifies temporal sequence and enforces a 5-second proximity constraint.

\paragraph{Final Blended Scoring Formula.}
Candidates are reranked by a deterministic blended score:
\begin{equation}
    \text{Score}(v) = S_{\text{summary}}(v) + 0.5 \cdot s(v, \mathcal{A}_q) + \Delta_{\text{verif}}(v)
\end{equation}
where $\Delta_{\text{verif}}(v)$ maps discrete verification outcomes:
\begin{equation}
    \Delta_{\text{verif}}(v) = \begin{cases}
        +6.0 & \text{if all verified predicates return } \texttt{TRUE}, \\
        -3.0 & \text{if any verified predicate returns } \texttt{FALSE}, \\
        \phantom{+}0.0 & \text{if verdicts are } \texttt{NULL}.
    \end{cases}
\end{equation}
A \texttt{NULL} verdict preserves the candidate's base score without adjustment, and ties are broken by initial dense retrieval order.

\paragraph{Max Mode Mechanics.}
Max Mode maximizes accuracy and recall via three mechanisms.
First, it widens $\text{DEP\_BAND}$ to $[0, 10]$ for candidates with confirmed anchors ($s(v, \mathcal{A}_q) \ge \tau_{\text{anchor}}$), disabling mid-band pruning to force exhaustive visual verification.
Second, rather than evaluating all video chunks in bulk, Max Mode queries the Judge LLM on each chunk output independently in parallel, preventing transient events from being diluted by neighboring chunks.
Third, a \textbf{Tier F3 Demotion-Repair Pass} inspects candidates from the top 3 of initial dense retrieval that were demoted without verification; if visual inspection confirms their anchor and dependents, they are restored to top ranks.

%% file: tex/evaluation.tex
\section{Evaluation} \label{sec:evaluation}
\subsection{Methodology}
\label{sec:eval-method}
\minihead{Implementation}
We implement \textsc{Vastra} in \texttt{Python 3.12} using an asynchronous, event-driven evaluation harness built with \texttt{asyncio}.
The evaluation harness runs the full online pipeline described in \S\ref{sec:system}: dense candidate retrieval with FAISS~\cite{faiss} over 384-dimensional \texttt{all-MiniLM-L6-v2}~\cite{wang2020minilm, reimers-2019-sentence-bert} embeddings of video summaries (retrieving top $N=40$ candidates); an LLM query router (\texttt{route\_query}) that decomposes each query into structured predicates and selects an execution route (\textsc{Route~A} summary scoring, \textsc{Route~B} visual clip examination, or \textsc{MIXED} hybrid verification); and a predicate-cascade agent (\texttt{run\_predicate\_cascade\_agent}) that issues \texttt{examine\_video\_footage} tool calls to a VLM with candidate short-circuiting.
Offline ingestion (transcript generation, embedding, and FAISS index construction) is computed once and reused across runs; only the online decision and verification path is exercised at evaluation time.
We build on \texttt{google-genai} and the OpenRouter API (\texttt{OpenRouterClientPool}) for model access.
All evaluation experiments run on a lightweight CPU virtual machine (AWS EC2 instance), leveraging cloud API calls for model inference.

\begin{table*}[t]
\centering
\caption{End-to-end performance comparison across 40 benchmark queries (mean $\pm$ std across three runs, $T=0.0$). \textbf{LLM Steps} measures reasoning and planning cycles, while \textbf{VLM Calls} measures multimodal video and audio inspections. \textsc{Vastra}'s presets establish distinct trade-offs across latency, cost, and determinism.}
\label{tab:end_to_end}
\small
\begin{tabular*}{\textwidth}{l @{\extracolsep{\fill}} ccccrrr}
\toprule
\textbf{System} & \textbf{$K$} & \textbf{mAP@20} & \textbf{Rec@20} & \textbf{Lat. (s)} & \textbf{Cost (\$)} & \textbf{LLM} & \textbf{VLM} \\
\midrule
\multicolumn{8}{l}{\textit{\textbf{Dense Vector Search}}} \\
FAISS Top-20 & 20 & $0.537 \pm 0.000$ & $0.834 \pm 0.000$ & \phantom{0}0.1 & $<0.0001$ & 0.0 & 0.0 \\
\midrule
\multicolumn{8}{l}{\textit{\textbf{Non-Agentic Baselines}}} \\
Pointwise-Summary & 20 & $0.553 \pm 0.000$ & $0.834 \pm 0.000$ & \phantom{0}2.5 & 0.0055 & 20.0 & 0.0 \\
Cascade-Summary@Top5 & 20 & $0.551 \pm 0.000$ & $0.834 \pm 0.000$ & \phantom{0}9.5 & 0.0141 & 1.0 & 5.0 \\
Cascade-Summary@Top10 & 20 & $0.554 \pm 0.000$ & $0.834 \pm 0.000$ & 11.2 & 0.0250 & 1.0 & 10.0 \\
Cascade-Dense@Top10 & 20 & $0.579 \pm 0.000$ & $0.834 \pm 0.000$ & 11.8 & 0.0223 & 0.0 & 10.0 \\
\midrule
\multicolumn{8}{l}{\textit{\textbf{ReAct Agents}}} \\
ReAct-S & 20 & $0.585 \pm 0.011$ & $0.834 \pm 0.000$ & 27.1 & 0.0115 & 1.0 & 0.0 \\
ReAct-S & 40 & $0.625 \pm 0.033$ & $0.837 \pm 0.002$ & 40.6 & 0.0199 & 1.0 & 0.0 \\
ReAct-S+E & 20 & $0.649 \pm 0.021$ & $0.834 \pm 0.000$ & 51.2 & 0.0458 & 3.4 & 5.8 \\
ReAct-S+E & 40 & $0.640 \pm 0.008$ & $0.824 \pm 0.005$ & 57.8 & 0.0597 & 3.2 & 6.6 \\
ReAct-IL & 20 & $0.551 \pm 0.011$ & $0.834 \pm 0.000$ & 81.0 & 0.1007 & 5.1 & 14.3 \\
ReAct-IL & 40 & $0.508 \pm 0.028$ & $0.802 \pm 0.021$ & 105.0 & 0.1350 & 5.0 & 21.3 \\
\midrule
\multicolumn{8}{l}{\textit{\textbf{\textsc{Vastra} Presets}}} \\
\textsc{Vastra}-Fast & 20 & $0.599 \pm 0.000$ & $0.834 \pm 0.000$ & \phantom{0}4.2 & 0.0044 & 2.0 & 0.0 \\
\textsc{Vastra}-Balanced & 20 & $0.608 \pm 0.000$ & $0.834 \pm 0.000$ & 22.1 & 0.0574 & 2.0 & 10.0 \\
\textsc{Vastra}-Max & 40 & $\mathbf{0.653} \pm 0.000$ & $\mathbf{0.846} \pm 0.000$ & 48.2 & 0.1022 & 2.0 & 20.9 \\
\bottomrule
\end{tabular*}
\end{table*}

\minihead{Models}
The system exercises two model roles: the \emph{decision LLM} (query routing, predicate decomposition, interval-summary scoring, and reranking decisions) and the \emph{verification VLM} (the engine behind \texttt{examine\_video\_footage} and \texttt{examine\_audio}).
Unless otherwise stated, both roles across \textsc{Vastra} and every ReAct baseline are served by \texttt{gemini-2.5-flash} at temperature $0$.\footnote{We compute all cost figures in this section at the model's list price at the time of our experiments: \$0.30 per 1M input tokens and \$2.50 per 1M output tokens, retrieved from the OpenRouter Gemini 2.5 Flash profile in August 2026.}
The reasoning (``thinking'') budget differs between architectures: ReAct baselines run with an $8192$-token thinking budget to support multi-step planning, whereas \textsc{Vastra} runs with the thinking budget disabled because the predicate cascade externalizes control flow into an explicit program.
The offline ingestion stack---Qwen-VL and Qwen-Omni for summary generation alongside \texttt{all-MiniLM-L6-v2} for embeddings---remains identical across all evaluated configurations.
Section~\ref{sec:eval-robustness} evaluates sensitivity across alternative model choices.

\minihead{Applications and datasets}
Our benchmark spans two video domains exercising distinct multimodal retrieval demands: indoor human activities from the Charades dataset~\cite{sigurdsson2016hollywood}, and dashcam and web footage collected from public videos covering driving incidents, acoustic cues, and long-form scenes.
The corpus contains 229 ingested videos, of which 145 are annotated as ground-truth targets.
Clip lengths range from $2.9$\,s to $240$\,s (mean $\approx 42$\,s), testing short-event localization and extended temporal reasoning.
We evaluate across 40 natural-language queries.
Relevance ground truth was established by dual independent human annotators evaluating candidate clips against explicit query criteria.
The 40 queries encompass 145 total relevant video occurrences (mean $3.6$ targets per query, ranging from 1 to 9).
Queries span 12 fine-grained categories across perceptual grounding (spatial, fine-grained, and audio-visual synchronization), temporal reasoning (sequence and persistence), and cognitive constraints (counting and cross-modal fusion).
Dense FAISS retrieval returns 40 candidate videos per query; systems rerank the top $K \in \{20, 40\}$ candidates while retaining the remaining candidates in their initial retrieval order.

\minihead{Baselines}
We implement all baselines inside the identical evaluation harness, sharing candidate retrieval pools ($K \in \{20, 40\}$), pre-computed video summaries, dense embeddings, and the \texttt{gemini-2.5-flash} backbone at temperature $T=0.0$.
We evaluate six baseline paradigms:
\begin{itemize}
    \item \textbf{Dense Retrieval Baseline (FAISS Top-$K$)}: Pure dense vector search over \texttt{all-MiniLM-L6-v2} embeddings of pre-ingested summaries directly through FAISS without downstream agentic reranking.
    \item \textbf{Pointwise Summary Scoring}: A simple non-decomposed baseline that evaluates candidate video summaries one at a time using \texttt{gemini-2.5-flash} ($T=0.0$) against the full natural language query, without query decomposition and without multimodal verification.
    \item \textbf{Non-Agentic Cascades}: Fixed retrieve-and-rerank pipelines that retrieve candidates from text summaries (or dense vectors) and inspect the top-$M$ clips ($M \in \{5, 10\}$) directly with the VLM without an iterative agent planning loop.
    \item \textbf{Summary-Only ReAct (ReAct-S)}: The agent receives pre-computed candidate video summaries directly within its prompt and produces a reranked list via LLM deliberation without issuing multimodal tool calls.
    \item \textbf{Multimodal Examination ReAct (ReAct-S+E)}: The agent receives candidate summaries and can invoke live multimodal verification tools (\texttt{examine\_video\_footage} and \texttt{examine\_audio}) on candidate intervals during each iteration before outputting its final ranking.
    \item \textbf{Interactive-Lookup ReAct (ReAct-IL)}: Modeled after video analytics architectures like Ava~\cite{ava}, ReAct-IL interactively queries indexed corpus metadata via 7 specialized tools (searching temporal objects, events, audio transcripts, and localized metadata) alongside direct VLM verification tools.
\end{itemize}

\paragraph{Working-Set Explosion in Interactive Video Search.}
While interactive knowledge-graph queries succeed in single-stream question answering~\cite{ava}, applying this pattern to Top-$K$ retrieval triggers a severe \textit{working-set explosion}.
Global search queries return every video matching a broad keyword, injecting dozens of extraneous candidates into the consideration pool.
Attempting to inspect these discovered videos through live VLM calls dilutes the prompt with noisy observations.
\textsc{Vastra} avoids working-set explosion by fixing the candidate pool upfront and restricting multimodal calls to candidate uncertainty bands.

\minihead{Metrics}
We measure ranking accuracy using Mean Average Precision at rank 20 (mAP@20)~\cite{manning2008introduction}.
For query $q$ with relevant set $R_q$,
\begin{equation}
\mathrm{AP@20}(q) = \frac{1}{\min(|R_q|, 20)} \sum_{k=1}^{20} P@k \cdot \mathrm{rel}_q(k),
\end{equation}
where $P@k$ is precision at rank $k$ and $\mathrm{rel}_q(k)$ indicates relevance.
We additionally report Recall@20 (the proportion of all target videos surfaced in the top 20).
Cost measures upstream LLM and VLM API inference spend per query in USD, parsed directly from provider response token telemetry.
Latency is end-to-end wall-clock time from query submission to final ranking, measured on an AWS EC2 c6i.2xlarge instance.
We execute each system configuration across three repeated runs and report mean $\pm$ sample standard deviation.

\subsection{End-to-end comparison to existing systems}
\label{sec:eval-e2e}

\minihead{Setup}
We evaluate \textsc{Vastra}'s three presets (Fast, Balanced, and Max) against dense retrieval, ReAct-S, ReAct-S+E, and ReAct-IL across all 40 benchmark queries under greedy decoding ($T=0.0$).

\minihead{Results and discussion}
Table~\ref{tab:end_to_end} establishes that \textsc{Vastra}'s presets deliver explicit operating trade-offs across latency, cost, and reproducibility:
\begin{itemize}[leftmargin=*]
    \item \textbf{\textsc{Vastra}-Fast} targets high-throughput triage by resolving queries entirely through parallel summary scoring ($0.0$ VLM calls). Compared to multimodal ReAct-S+E ($K=20$), it provides a $12.2\times$ latency reduction ($4.2$\,s vs.\ $51.2$\,s) and $10.4\times$ cost reduction ($\$0.0044$ vs.\ $\$0.0458$) at a $5.0$-point mAP drop ($0.599$ vs.\ $0.649$). Against summary-only ReAct-S ($K=20$), \textsc{Vastra}-Fast yields $+1.4$ higher mAP points while executing $6.5\times$ faster and costing $2.6\times$ less, without run-to-run variance.
    \item \textbf{\textsc{Vastra}-Balanced} introduces selective visual clip inspection for candidates with uncertain summaries, cutting latency by $2.3\times$ ($22.1$\,s vs.\ $51.2$\,s) relative to ReAct-S+E ($K=20$) while outperforming interactive ReAct-IL by $+10.0$ mAP points at $4.8\times$ lower latency and $2.4\times$ lower spend.
    \item \textbf{\textsc{Vastra}-Max} outperformes peak baseline retrieval accuracy ($0.653$ vs.\ $0.649$ mAP) at comparable latency ($48.2$\,s vs.\ $51.2$\,s) by expanding candidate evaluation to $K=40$. Rather than minimizing spend, \textsc{Vastra}-Max trades additional verification calls on double size candidate list for exact run-to-run determinism under greedy serving, eliminating the rank variance observed in ReAct baselines ($\sigma = \pm 0.021$).
\end{itemize}

\paragraph{Comparison to Non-Agentic and Pointwise Baselines.}
To isolate whether agentic decomposition and verification are necessary, we compare against simpler non-agentic alternatives: pointwise summary scoring (evaluating candidate summaries individually against the raw query without decomposition or verification) and two-stage cascades (retrieving from summaries or dense vectors and reranking the top-$M$ candidates with a VLM without an agent loop).
As recorded in Table~\ref{tab:end_to_end}, scoring candidate summaries one at a time (\texttt{Pointwise-Summary}) achieves $0.553$ mAP@20 at $2.5$\,s latency and $\$0.0055$ cost ($0.0$ VLM calls).
While fast and inexpensive, it trails listwise ReAct-S ($0.585$) by $3.2$ mAP points due to score calibration drift across independently evaluated summaries.
Crucially, \textsc{Vastra}-Fast also issues zero visual verification calls ($0.0$ VLM), but achieves $0.599$ mAP@20---surpassing \texttt{Pointwise-Summary} by $+4.6$ points at comparable latency ($4.2$\,s vs.\ $2.5$\,s) and lower spend ($\$0.0044$ vs.\ $\$0.0055$).
This gap demonstrates the empirical value of query routing and predicate decomposition: evaluating structured atomic clauses against interval summaries extracts decisive signals that monolithic query scoring overlooks.
Adding selective VLM inspection to non-agentic candidates yields modest improvements ($0.554$ for \texttt{Cascade-Summary@Top10} and $0.579$ for \texttt{Cascade-Dense@Top10}), but remains below \textsc{Vastra}-Balanced ($0.608$, $+2.9$ to $+5.4$ points).
Static top-$M$ cascades suffer from blind truncation: when first-stage text summaries miss subtle visual or acoustic cues, relevant videos drop to ranks 11--20 where the cascade never inspects them.
\textsc{Vastra}'s uncertainty-guided tiering routes candidates dynamically based on interval-summary score ambiguity ($\tau_{\text{low}} \le s \le \tau_{\text{high}}$), targeting inspection where confidence is intermediate rather than expending budget on already-certain top candidates.

\paragraph{Category Breakdown Across Query Archetypes.}
Evaluating query categories confirms that topological routing aligns verification spend with query complexity (\S\ref{sec:motivation-decomposition}).
For single visual entities and explicit events, \textsc{Route~A} resolves candidates directly from summaries with zero visual inspection calls.
For multi-step temporal sequences, persistence constraints, and fine-grained audio-visual synchronization, escalating candidates to \textsc{Route~B} or \textsc{Route~MIXED} yields $+28$ to $+38$ AP points over monolithic prompts, while bypassing decomposition on atomic actions prevents false-negative cascade demotions.

\begin{figure*}[t]
\centering
\includegraphics[width=\textwidth]{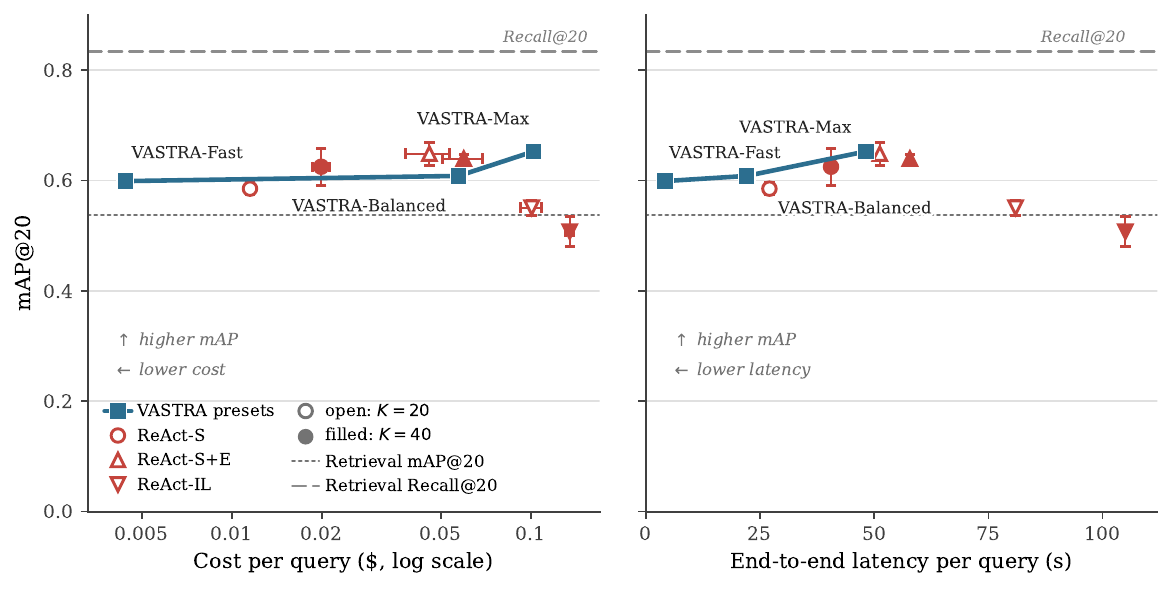}
\caption{Accuracy vs.\ per-query cost (left; log scale) and latency (right). Filled blue squares represent \textsc{Vastra}'s presets (Fast, Balanced, Max; Table~\ref{tab:end_to_end}). Red markers show ReAct baselines across candidate pool sweeps (open $K=20$, filled $K=40$), with error bars indicating $\pm 1$ standard deviation over three runs. The lower dotted line indicates first-stage retrieval mAP@20; the upper dashed line marks first-stage Recall@20.}
\label{fig:pareto}
\end{figure*}

\subsection{Consistency and determinism}
\label{sec:eval-determinism}

\minihead{Setup}
Production video search requires reproducible rankings across repeated submissions against a static corpus.
We evaluate consistency over three independent runs across all 40 queries under greedy decoding ($T=0.0$), measuring rank stability via the pairwise Kendall rank correlation coefficient ($\tau$) alongside accuracy variance ($\sigma_{\text{mAP}}$), latency, and cost.
To isolate the origin of agentic variance, we execute a thinking budget isolation sweep across $B_{\text{think}} \in \{0, 4096, 8192\}$ tokens for \textsc{Vastra} and ReAct-S+E with prompt caching disabled (Table~\ref{tab:determinism}).

\begin{table}[t]
\centering
\caption{Determinism and rank stability across three repeated runs on 40 benchmark queries ($T=0.0$). \textsc{Vastra} achieves invariant control flow ($\tau = 1.000$) whereas unconstrained reasoning degrades stability.}
\label{tab:determinism}
\scriptsize
\setlength{\tabcolsep}{1.6pt}
\begin{tabular*}{\columnwidth}{l @{\extracolsep{\fill}} ccccrrr}
\toprule
\textbf{System} & \textbf{$B_{\text{th}}$} & \textbf{$K$} & \textbf{mAP@20} & \textbf{$\tau$} & \textbf{Lat.} & \textbf{Cost} & \textbf{LLM/VLM} \\
\midrule
\multicolumn{8}{l}{\textit{\textbf{\textsc{Vastra}}}}\\
Fast & 0 & 20 & $0.599 \pm 0.000$ & $\mathbf{1.000}$ & \phantom{0}4.2s & \$0.004 & 2.0 / 0.0 \\
Balanced & 0 & 20 & $0.608 \pm 0.000$ & $\mathbf{1.000}$ & 22.1s & \$0.057 & 2.0 / 10.0 \\
Max & 0 & 40 & $\mathbf{0.653} \pm 0.000$ & $\mathbf{1.000}$ & 48.2s & \$0.102 & 2.0 / 20.9 \\
\midrule
\multicolumn{8}{l}{\textit{\textbf{ReAct Baselines}}}\\
ReAct-S & 8k & 40 & $0.625 \pm 0.033$ & 0.918 & 40.6s & \$0.020 & 1.0 / 0.0 \\
ReAct-S+E & 8k & 20 & $0.649 \pm 0.021$ & 0.758 & 51.2s & \$0.046 & 3.4 / 5.8 \\
ReAct-S+E & 8k & 40 & $0.640 \pm 0.008$ & 0.789 & 57.8s & \$0.060 & 3.2 / 6.6 \\
ReAct-IL & 8k & 40 & $0.508 \pm 0.028$ & 0.761 & 105s & \$0.135 & 5.0 / 21.3 \\
\midrule
\multicolumn{8}{l}{\textit{\textbf{Thinking Budget Sweep}}}\\
ReAct-S+E & 0 & 20 & $0.599 \pm 0.037$ & 0.688 & 49.0s & \$0.058 & 3.8 / 15.0 \\
\textsc{Vastra}-Bal. & 4k & 20 & $0.534 \pm 0.025$ & 0.676 & 57.4s & \$0.050 & 2.0 / 10.0 \\
\textsc{Vastra}-Bal. & 8k & 20 & $0.521 \pm 0.017$ & 0.719 & 39.2s & \$0.042 & 2.0 / 8.7 \\
\bottomrule
\end{tabular*}
\end{table}

\minihead{Results and discussion}
Table~\ref{tab:determinism} demonstrates that agentic ranking instability originates from open-ended control flow rather than reasoning scratchpads:
\begin{itemize}[leftmargin=*]
    \item \textbf{Ablating thinking tokens fails to stabilize ReAct.}
    Zeroing the reasoning budget ($B_{\text{think}}=0$) on ReAct-S+E reduces latency slightly ($51.2$\,s to $49.0$\,s), but drops rank stability to $\tau = 0.688$ and widens accuracy variance to $\sigma = \pm 0.037$, lowering mAP from $0.649$ to $0.599$.
    Without intermediate scratchpads, decision entropy increases, driving VLM tool calls from $5.8$ to $15.0$ and cost to $\$0.0581$.
    ReAct's variance is structural: open conversational loops diverge across runs even without thinking tokens.
    \item \textbf{Deterministic control flow in \textsc{Vastra}.}
    \textsc{Vastra} guarantees reproducible rankings by compiling queries into declarative predicate graphs.
    Because verification employs zero temperature, discrete three-valued verdicts, and deterministic score aggregation, execution control flow remains invariant across runs.
    Tool invocations remain bounded across presets ($0.0$, $10.0$, and $20.9$ calls), contrasting with ReAct's open-ended horizons.
    \item \textbf{Negative impact of latent reasoning on structured pipelines.}
    Forcing \textsc{Vastra} to generate extended reasoning tokens ($B=4096, 8192$) lowers mAP to $0.534$ and $0.521$ while reducing rank stability ($\tau = 0.676$ and $0.719$).
    Verbose chain-of-thought injects hallucinations into narrow perceptual checks and causes premature candidate short-circuiting during filtering.
    Operating at $B_{\text{think}}=0$ is the natural deployment configuration unlocked by predicate compilation, securing deterministic rankings without reasoning overhead.
\end{itemize}

\subsection{Sensitivity to VLM and Decision LLM models}
\label{sec:eval-robustness}

\begin{figure}[t]
\centering
\makebox[\columnwidth][c]{%
  \includegraphics[trim=5pt 10pt 17pt 14pt, clip, width=1.08\columnwidth]{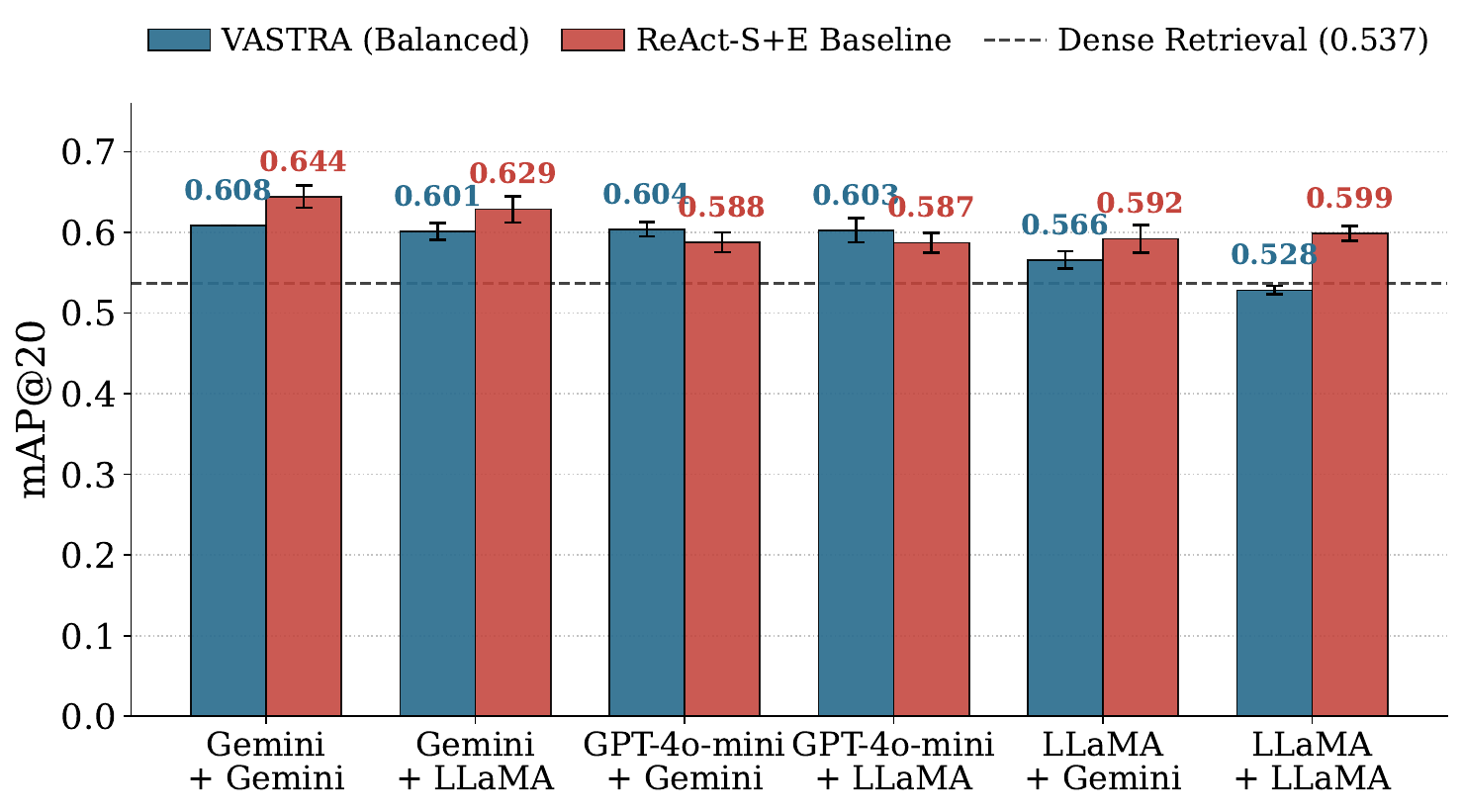}%
}
\vspace{-2pt} %
\caption{\textbf{Model robustness across decision-LLM $\times$ VLM pairings} ($\pm$ std). The dashed line marks first-stage dense retrieval ($0.537$). While ReAct-S+E (red) drops by $5.6$ points under \texttt{gpt-4o-mini}, \textsc{Vastra}-Balanced (blue) remains stable, degrading by only $0.4$ points.}
\label{fig:robustness}
\end{figure}

\minihead{Setup}
We independently vary the decision LLM and verification VLM across \texttt{gemini-2.5-flash}, \texttt{gpt-4o-mini}, and \texttt{llama-4-maverick} for both \textsc{Vastra} (Balanced) and the multimodal \mbox{ReAct-S+E} baseline ($K=20$), holding prompts, candidate pools, and greedy decoding fixed.
Each of the 6 combinations is evaluated across repeated runs (Table~\ref{tab:robustness} and Figure~\ref{fig:robustness}).

\begin{table}[t]
\centering
\caption{Model robustness across decision-LLM $\times$ verification-VLM swaps for \textsc{Vastra} (Balanced) and the multimodal \mbox{ReAct-S+E} baseline ($K=20$, mean $\pm$ std across repeated runs). gem = \texttt{gemini-2.5-flash}, gpt = \texttt{gpt-4o-mini}, llama = \texttt{llama-4-maverick}.}
\label{tab:robustness}
\setlength{\tabcolsep}{2.5pt}
\resizebox{\columnwidth}{!}{%
\begin{tabular}{llcccc}
\toprule
& & \multicolumn{2}{c}{\textbf{\textsc{Vastra}-Balanced}} & \multicolumn{2}{c}{\textbf{ReAct-S+E Baseline}} \\
\cmidrule(lr){3-4} \cmidrule(lr){5-6}
\textbf{LLM} & \textbf{VLM} & \textbf{mAP@20} & \textbf{Lat. (s)} & \textbf{mAP@20} & \textbf{Lat. (s)} \\
\midrule
gem   & gem   & $0.608 \pm 0.000$ & $22.1$ & $0.644 \pm 0.014$ & $54.5$ \\
gpt   & gem   & $0.604 \pm 0.009$ & $33.4$ & $0.588 \pm 0.012$ & $22.8$ \\
gem   & llama & $0.601 \pm 0.010$ & $38.9$ & $0.629 \pm 0.016$ & $47.7$ \\
gpt   & llama & $0.603 \pm 0.015$ & $47.7$ & $0.587 \pm 0.012$ & $33.7$ \\
llama & gem   & $0.566 \pm 0.011$ & $58.1$ & $0.592 \pm 0.017$ & $56.4$ \\
llama & llama & $0.528 \pm 0.005$ & $62.4$ & $0.599 \pm 0.009$ & $58.0$ \\
\bottomrule
\end{tabular}%
}
\end{table}

\minihead{Results and discussion}
Comparing \textsc{Vastra} and ReAct-S+E under model swaps reveals fundamental architectural differences in model sensitivity (Table~\ref{tab:robustness} and Figure~\ref{fig:robustness}):
\begin{itemize}[leftmargin=*]
    \item \textbf{Autonomous planning degrades sharply under lightweight decision LLMs.}
    Swapping the decision LLM to \texttt{gpt-4o-mini} causes ReAct-S+E's retrieval accuracy to plunge by $5.6$ mAP points ($0.644 \to 0.588$ with Gemini VLM; $0.629 \to 0.587$ with LLaMA VLM).
    Unconstrained ReAct execution places heavy demands on the LLM's multi-step planning, working memory, and tool-invocation syntax; lightweight models struggle to maintain coherent search trajectories across iterative tool turns.
    In contrast, \textsc{Vastra} externalizes control flow into an explicit cascade and topological router.
    Under \texttt{gpt-4o-mini}, \textsc{Vastra} drops by a negligible $0.004$ mAP points ($0.608 \to 0.604$), decisively outperforming ReAct-S+E ($0.604$ vs.\ $0.588$) while remaining above $0.60$ mAP even when paired with an open-weights VLM ($0.603$).
    \item \textbf{Verification VLM sensitivity is modest across both paradigms.}
    Replacing \texttt{gemini-2.5-flash} with \texttt{llama-4-maverick} for video verification causes only minor degradation across both systems ($0.608 \to 0.601$ for \textsc{Vastra}; $0.644 \to 0.629$ for ReAct).
    Because candidate intervals and questions are scoped to concrete perceptual queries, open-weights VLMs deliver sufficient discriminative power.
    \item \textbf{Open-weights decision LLMs.}
    Under the open-weights \texttt{llama-4-maverick} decision LLM, both systems experience performance drops ($0.566$ for \textsc{Vastra}; $0.592$ for ReAct).
    For \textsc{Vastra}, the drop concentrates in the query routing stage where complex multi-predicate decomposition demands frontier instruction following; dual-role \texttt{llama} represents the only configuration dipping below the dense retrieval baseline ($0.528$).
\end{itemize}

\subsection{Ablation of \sn's components}
\label{sec:eval-ablation}

\minihead{Setup}
We ablate each pipeline component from the default full configuration ($K=20$, $0.6216$ mAP@20, $\$0.0693$, $30.9$\,s, $13.3$ VLM calls) across all 40 queries (Table~\ref{tab:ablation}).
This ablation baseline operates with an unconstrained verification budget to isolate the impact of each architectural mechanism without verification truncation; in contrast, the deployable \textsc{Vastra}-Balanced preset (\S\ref{sec:eval-e2e}) enforces a strict verification cap ($10$ calls), completing faster and cheaper ($22.1$\,s, $\$0.0574$, $0.608$ mAP) while retaining $98\%$ of peak accuracy.

\begin{table}[t]
\centering
\caption{\textbf{Ablation of \textsc{Vastra} components.} Each row disables or alters one component from the default system at $K=20$ (top row); the bottom row evaluates candidate set expansion ($K=40$).}
\label{tab:ablation}
\footnotesize
\setlength{\tabcolsep}{2pt}
\begin{tabular*}{\columnwidth}{l @{\extracolsep{\fill}} cccc}
\toprule
\textbf{Configuration} & \textbf{mAP@20} & \textbf{Cost (\$)} & \textbf{Lat.\ (s)} & \textbf{VLM} \\
\midrule
\textsc{Vastra} (full default) & $\mathbf{0.6216}$ & $0.0693$ & $30.9$ & $13.3$ \\
\midrule
w/o router (always Route~B) & $0.5816$ & $0.0737$ & $13.1$ & $14.8$ \\
w/o router (always Route~A) & $0.5990$ & $0.0044$ & $\phantom{0}4.2$ & $\phantom{0}0.0$ \\
w/o predicate decomp. & $0.5620$ & $0.0582$ & $14.5$ & $11.3$ \\
w/o short-circuiting & $0.5697$ & $0.0448$ & $10.8$ & $\phantom{0}8.5$ \\
w/o mid-band filtering & $0.5797$ & $0.0656$ & $15.6$ & $12.5$ \\
w/o fallback tiers & $0.5491$ & $0.0115$ & $\phantom{0}3.2$ & $\phantom{0}1.9$ \\
w/o 3-valued verification & $0.5809$ & $0.0620$ & $17.5$ & $11.8$ \\
\midrule
candidate set $K{=}40$ & $0.6528$ & $0.1022$ & $48.2$ & $20.9$ \\
\bottomrule
\end{tabular*}
\end{table}

\minihead{Results and discussion}
Disabling individual components lowers mAP@20 by $2.3$ to $7.3$ points, revealing distinct architectural roles:
\begin{itemize}[leftmargin=*]
    \item \textbf{Fallback tiers are critical to recall.}
    Disabling Fallback tiers causes the largest accuracy regression ($-7.3$ points to $0.5491$) while collapsing latency and cost ($3.2$\,s, $\$0.0115$).
    Without fallback escalation, candidates with unresolvable predicates are abandoned, causing premature termination on complex queries.
    \item \textbf{Structured decomposition outperforms monolithic verification.}
    Replacing structured predicates with monolithic prompts drops mAP by $6.0$ points ($0.5620$) with minor spend reduction ($11.3$ calls, $\$0.0582$), confirming that verifying decomposed sub-clauses yields higher precision than prompting against whole queries.
    \item \textbf{Query routing reduces cost while improving accuracy.}
    Forcing all queries to Route~B lowers accuracy by $4.0$ points while increasing cost ($0.5816$ mAP at $\$0.0737$).
    Conversely, Route~A achieves $0.5990$ mAP at minimal cost ($\$0.0044$).
    Candidate short-circuiting, mid-band filtering, and three-valued verification spend targeted inspection budget to protect ranking accuracy against ambiguous classifications.
    \item \textbf{Candidate-set expansion scales precision.}
    Expanding the candidate pool to $K=40$ raises mAP@20 to $0.6528$ (+3.1 points) by surfacing relevant videos from ranks 21--40, scaling latency and spend predictably.
\end{itemize}

%% file: tex/related_work.tex
\section{Related Work} \label{sec:related_work}

\subsection{Classical Video Analytics Systems}
Prior to foundation models, video analytics systems accelerated queries by bypassing expensive neural networks on uninformative frames.
One dominant approach deploys filter-and-refine cascades, placing specialized lightweight proxies ahead of costly reference models~\cite{noscope,blazeit,focus,boggart,videochef,everest,eva,seiden}.
To support richer queries, multi-model query planners optimize declarative operator graphs over fixed model catalogs~\cite{viva,figo}, while adaptive frameworks retrain models continually to mitigate real-world distribution shift~\cite{vulcan,remembrall,ekya,recl,caravan}.
Other systems explore interactive, exploratory video search interfaces~\cite{vocalexplore}.
While these systems bound latency and cost, they require predefined schema targets and cannot evaluate free-form natural-language queries.

\subsection{VLM-Based Video Analytics and Retrieval}
Vision-language models eliminate fixed-predicate constraints by mapping video content and natural language into shared semantic spaces.
One class of systems employs multimodal embeddings to support similarity search and retrieval-augmented generation~\cite{clip,blip2,flamingo,zelda,ava,lotus}.
Another class fine-tunes multimodal language models to generate open-ended answers directly from video sequences across diverse temporal horizons~\cite{videochatgpt,videochat,timechat,moviechat,llava_video}.
However, embedding-based RAG struggles to capture fine-grained temporal and causal constraints, while end-to-end video LLMs incur prohibitive computational costs when evaluated across candidate pools.

\subsection{Interactive Video Agents}
Interactive agent architectures interleave reasoning steps with tool execution, using intermediate observations to guide downstream actions~\cite{yao2023react}.
In video domains, agents dynamically query computer vision detectors, acoustic classifiers, and specialized metadata indices to analyze long-form footage~\cite{yang2023mmreact,fan2025videoagent,gao2023assistgpt,zhang2024omagent} or rerank video candidates~\cite{skow2026rankvideoreasoningrerankingtexttovideo}.
Unlike \textsc{Vastra}, existing video agents rely on unconstrained deliberation loops that produce non-deterministic rankings, suffer from working-set explosion, and incur unbounded latency across multi-candidate retrieval workloads.

%% file: tex/conclusion.tex
\section{Conclusion and Future Directions}
\label{sec:conclusion}

\textsc{Vastra} treats agentic video retrieval as a systems problem rather than an unconstrained prompting task.
Its core architectural design externalizes control flow from the model into a deterministic predicate execution graph, scoring answerable queries directly from cached text summaries and restricting expensive VLM inspections strictly to uncertain candidates.
On our benchmark, \textsc{Vastra}-Fast provides a $12.2\times$ latency reduction and $10.4\times$ cost reduction relative to multimodal ReAct ($4.2$\,s, $\$0.0044$), while outperforming summary-only ReAct-S by $+1.4$ mAP points at $6.5\times$ speedup.
At the high-accuracy tier, \textsc{Vastra}-Max matches peak baseline accuracy ($0.653$ vs.\ $0.649$ mAP) while delivering exact run-to-run determinism under greedy serving, eliminating the rank variance inherent in conversational agent loops.

\paragraph{Self-improving summaries.}
Targeted verification verdicts represent high-confidence query observations that are discarded after ranking.
Writing these verified predicates back into cached video summaries would enable future queries sharing related attributes to resolve directly on the summary path, progressively lowering verification costs across repeated workloads.

\paragraph{Budget- and SLO-driven execution.}
While \textsc{Vastra} currently exposes discrete operating presets, its predictable cost model supports planning execution directly against user-specified latency and cost budgets~\cite{circinus}.
Future execution engines can dynamically adapt candidate admission thresholds and verification depth to maximize retrieval precision under fixed service-level objectives.

\paragraph{Scaling to larger candidate sets and corpora.}
Our benchmark evaluates 40 queries across a curated corpus of 229 videos, demonstrating the mechanics and efficiency gains of deterministic predicate cascades over open-ended ReAct loops.
However, this scope provides limited empirical evidence of how the pipeline scales to massive video collections with thousands of hours of footage or initial candidate sets orders of magnitude larger ($K \gg 40$).
Future work will examine hierarchical candidate pruning, distributed predicate verification, and indexing optimizations required to maintain strict latency and cost bounds on enterprise-scale video repositories.